\documentclass[twocolumn]{aastex631}

\usepackage{ulem}
\usepackage{natbib}
\usepackage{amsmath}

\usepackage{color}
\newcommand{\fdir}{./figure}
\newcommand{\fmastro}{f_{\rm m,astro}}
\newcommand{\fmspectro}{f_{\rm m,spectro}}
\newcommand{\fmspectroconv}{\hat{f}_{\rm m,spectro}}
\newcommand{\msun}{M_\odot}

\begin{document}

\title{Search for a Black Hole Binary in Gaia DR3 Astrometric Binary
  Stars with Spectroscopic Data}

\correspondingauthor{Ataru Tanikawa}
\email{tanikawa@ea.c.u-tokyo.ac.jp}

\author[0000-0002-8461-5517]{Ataru Tanikawa}
\affiliation{Department of Earth Science and Astronomy, College of
  Arts and Sciences, The University of Tokyo, 3-8-1 Komaba, Meguro-ku,
  Tokyo 153-8902, Japan}

\author[0000-0001-6924-8862]{Kohei~Hattori}
\affiliation{National Astronomical Observatory of Japan, 2-21-1 Osawa, Mitaka, Tokyo 181-8588, Japan}
\affiliation{Institute of Statistical Mathematics, 10-3 Midoricho, Tachikawa, Tokyo 190-8562, Japan}

\author[0000-0001-8181-7511]{Norita~Kawanaka}
\affiliation{Center for Gravitational Physics and Quantum Information,
  Yukawa Institute for Theoretical Physics, Kyoto University,
  Kitashirakawa Oiwake-cho, Sakyo-ku, Kyoto 606-8502, Japan}
  
\author[0000-0002-3033-4576]{Tomoya~Kinugawa}
\affiliation{Institute for Cosmic Ray Research, University of Tokyo, Kashiwa, Chiba 255-8582, Japan}

\author[0000-0002-3561-8658]{Minori~Shikauchi}
\affiliation{Department of Physics, the University of Tokyo, 
7-3-1 Hongo, Bunkyo, Tokyo 113-0033, Japan}
\affiliation{Research Center for the Early Universe (RESCEU), School
  of Science, The University of Tokyo, 7-3-1 Hongo, Bunkyo-ku, Tokyo
  113-0033, Japan}
\affiliation{Department of Physics and Astronomy, the University of British Columbia, 6224 Agricultural Road, Vancouver, BC, V6T 1Z1, Canada}

\author[0000-0002-6347-3089]{Daichi~Tsuna}
\affiliation{Research Center for the Early Universe (RESCEU), School
  of Science, The University of Tokyo, 7-3-1 Hongo, Bunkyo-ku, Tokyo
  113-0033, Japan}

\begin{abstract}

  We report the discovery of a candidate binary system consisting of a
  black hole (BH) and a red giant branch star from the Gaia DR3. This
  binary system is discovered from 64108 binary solutions for which
  both astrometric and spectroscopic data are available. For this
  system, the astrometric and spectroscopic solutions are consistent
  with each other, making this system a confident candidate of a BH
  binary. The primary (visible) star in this system, Gaia DR3
  5870569352746779008, is a red giant branch whose mass is quite
  uncertain. Fortunately, albeit the uncertainty of the primary's
  mass, we can estimate the mass of the secondary (dark) object in
  this system to be $>5.68$ $M_\odot$ with a probability of $99$ \%,
  based on the orbital parameters. The mass of the secondary object is
  much larger than the maximum neutron star mass ($\sim 2.0$
  $M_\odot$), which indicates that the secondary object is likely a
  BH. We argue that, if this dark object is not a BH, this system must
  be a more exotic system, in which the primary red giant branch star
  orbits around a quadruple star system (or a higher-order multiple
  star system) whose total mass is more than $5.68$ $M_\odot$. If this
  is a genuine BH binary, this has the longest period ($1352.22 \pm
  45.81$ days) among discovered so far. As our conclusion entirely
  relies on the Gaia DR3 data, independent confirmation with follow-up
  observations (e.g. long-term spectra) is desired.

\end{abstract}


\section{Introduction}
\label{sec:Introduction}

Stellar-mass black holes (BHs) are the final state of massive stars
with several $10$ $\msun$ \citep[e.g.][]{2002RvMP...74.1015W}. BHs are
not just dark, especially when they are members of close binary
stars. Thus, they have been discovered as X-ray binaries
\citep[e.g.][]{2017hsn..book.1499C} and gravitational wave transients
\citep{2019PhRvX...9c1040A, 2021PhRvX..11b1053A,
  2021arXiv211103634T}. Nevertheless, since such BH populations are
rare, just a handful of BHs are known. So far, $\sim 100$ BHs have
been detected as X-ray binaries in the Milky Way,
\citep{2016A&A...587A..61C}, while there should be $\sim 10^8$ BHs in
the Milky Way \citep[e.g.][]{1983bhwd.book.....S,
  1992A&A...262...97V}. This is because BHs are bright in X-rays only
when they have close companion stars: binary periods of less than
about $1$ day.

Great efforts have been made to discover a variety of BHs in binary
stars (hereafter BH binaries). Many spectroscopic observations have
reported BH binaries with periods of $1$--$100$ days
\citep{2019Sci...366..637T, 2019Natur.575..618L, 2020A&A...637L...3R,
  2021MNRAS.504.2577J, 2022MNRAS.tmp.2111J, 2022A&A...665A.180L,
  2022MNRAS.511.2914S}. However, many concerns have been raised for
these reports \citep{2020Natur.580E..11A, 2020MNRAS.493L..22E,
  2020Sci...368.3282V, 2021MNRAS.502.3436E, 2020MNRAS.495.2786E,
  2020A&A...633L...5I, 2020PASJ...72...39T, 2020ApJ...901..116S,
  2020A&A...641A..43B, 2020A&A...639L...6S, 2022MNRAS.511L..24E,
  2022MNRAS.511.3089E, 2022MNRAS.512.5620E}.  Several BH binaries
\citep{2018MNRAS.475L..15G, 2022NatAs...6.1085S} still survive,
despite such harsh environment for BH binary searchers.

{\it Gaia} have monitored more than $10^9$ stars and their astrometric
and spectroscopic motions during 34 months \citep{2016A&A...595A...1G,
  2018A&A...616A...1G, 2021A&A...650C...3G, 2022arXiv220800211G}, and
have published $\sim 3 \times 10^5$ astrometric and spectroscopic
binary stars in total in Gaia Data Release 3
\citep[GDR3;][]{2022arXiv220605595G, 2022arXiv220605439H,
  2022arXiv221211971H, 2022arXiv220605726H}. Before GDR3, many studies
have predicted that {\it Gaia} discovers a large amount of compact
objects in binary stars, such as white dwarfs (WDs), neutron stars
(NSs), and BHs, from {\it Gaia}'s astrometric data
\citep{2017MNRAS.470.2611M, 2017ApJ...850L..13B, 2018MNRAS.481..930Y,
  2018ApJ...861...21Y, 2018arXiv181009721K, 2019MNRAS.487.5610S,
  2019ApJ...885..151S, 2019ApJ...886...68A, 2021arXiv211005549A,
  2020PASJ...72...45S, 2022ApJ...928...13S, 2022ApJ...931..107C,
  2022A&A...658A.129J}. Starting with \cite{2022arXiv220605595G}, many
research groups have searched for WD, NS, and BH binaries in
spectroscopic binaries \citep{2022arXiv220606032G,
  2022arXiv220705086J, 2022ApJ...940..126F} and astrometric binaries
\citep{2022arXiv220700680A, 2022arXiv221005003C, 2023MNRAS.518.1057E,
  2023MNRAS.518.2991S} just after GDR3.

GDR3 has presented several $10^4$ binary stars with both of
astrometric and spectroscopic data. However, previous studies have
focused on either of astrometric or spectroscopic data. In this paper,
we first search for BH binaries from binary stars where both data are
available, taking into account both of astrometric and spectroscopic
data. In other words, we first make a comparison between astrometric
and spectroscopic mass functions (see Eqs. (\ref{eq:fmastro1}) and
(\ref{eq:fmspectro1}), respectively) to search for BH
binaries.

We eventually find a promising BH binary candidate whose source ID is
GDR3 5870569352746779008. After we posted this work to arXiv,
\cite{2023MNRAS.518.1057E} have independently pointed out that the BH
binary candidate is promising, and \cite{2023arXiv230207880E} have
confirmed it as a genuine BH binary by follow-up observations. This
shows that our search is helpful and efficient to narrow down BH
binary candidates. Although we recognize that
\cite{2023arXiv230207880E} call it ``Gaia BH2'', we call it ``BH
binary candidate'' in this paper. This is {\it not} because we
disagree with their confirmation, but because we regarded it as a BH
binary candidate when we posted this work to arXiv (September 2022).

The structure of this paper is as follows. In section
\ref{sec:SampleSelection}, we describe how to select a sample of
binary stars from GDR3, and how to list up BH binary
candidates. Finally, we find one BH binary candidate. In section
\ref{sec:Analysis}, we analyze the BH binary candidate in detail. In
section \ref{sec:Discussion}, we discuss the BH binary candidate,
comparing it with BH binary candidates listed by previous studies. In
section \ref{sec:Summary}, we summarize this paper.
 
\section{Sample selection}
\label{sec:SampleSelection}

\subsection{Search for BH binaries with $m_2 > 3 \msun$}
\label{sec:SearchForBH}

We select GDR3 binary stars with astrometric and spectroscopic data
\citep{2022arXiv220605595G}. There are three types of such binary
stars. The orbital solutions of the first type are obtained from
astrometric and spectroscopic data. They have a {\tt
  nss\_solution\_type} name of ``{\tt AstroSpectroSB1}'' in the
non-single star tables of GDR3 ({\tt nss\_two\_body\_orbit}). We call
them {\tt AstroSpectroSB1} binary stars. The second type has an
orbital solution derived only from astrometric data, and additionally
has the total amplitude in the radial velocity time series called
``{\tt rv\_amplitude\_robust}''. Such binary stars have a {\tt
  nss\_solution\_type} name of ``{\tt Orbital}'', and satisfy the
following two conditions. First, they are bright stars; they have {\it
  Gaia} RVS magnitude less than and equal to 12. Second, their radial
velocities are computed more than twice. For the third type, binary
stars have two {\tt nss\_solution\_type} names of ``{\tt Orbital}''
and ``{\tt SB1}'' independently. Such binary stars also have a {\tt
  non\_single\_star} value of 3. Hereafter, the second and third types
are collectively called {\tt Orbital} binary stars simply. We can
extract such a sample of binary stars from GDR3 with following ADQL
query:
\begin{verbatim}
select nss.*, gs.*
from gaiadr3.nss_two_body_orbit as nss,
gaiadr3.gaia_source as gs
where nss.source_id = gs.source_id
and (nss.nss_solution_type = 'AstroSpectroSB1'
     or (nss.nss_solution_type = 'Orbital'
         and gs.rv_amplitude_robust IS NOT NULL)
     or gs.non_single_star = 3)
\end{verbatim}
The line numbers 5, 6-7, and 8 in the above ADQL query try to pick up
the first, second and third types, respectively. However, the line
number 8 picks up binary stars not only the third type binaries but
also many other binaries, for example, binary stars with {\tt
  nss\_solution\_type} names of ``{\tt acceleration7}'' and ``{\tt
  SB1}''. We exclude them later. Finally, the total number of binary
stars is 64108 consisting of 33467 ``AstroSpectroSB1'' and 30641
``Orbital'' binary stars, where the numbers of the second and third
types are 30629 and 12, respectively.

We search for BH binary candidates from the above sample, using
astrometric and spectroscopic mass functions ($\fmastro$ and
$\fmspectro$, respectively). We express these mass functions as
follows:
\begin{align}
  \fmastro &= (m_1+m_2) \left| \frac{m_2}{m_1+m_2} -
    \frac{F_2/F_1}{1+F_2/F_1} \right|^3 \label{eq:fmastro1} \\
  &= 1 \left( \frac{a_1}{{\rm mas}} \right)^3 \left(
  \frac{\varpi}{{\rm mas}} \right)^{-3} \left( \frac{P}{{\rm yr}}
  \right)^{-2} \; [\msun], \label{eq:fmastro2}
\end{align}
and
\begin{align}
  \fmspectro &= (m_1+m_2) \left( \frac{m_2}{m_1+m_2}
  \right)^3 \label{eq:fmspectro1} \\
  &= 3.7931 \times 10^{-5} \left( \frac{K_1}{{\rm km~s^{-1}}}
  \right)^3 \left( \frac{P}{{\rm yr}} \right) \nonumber \\
  &\times (1-e^2)^{3/2} \sin^{-3} i \; [\msun], \label{eq:fmspectro2}
\end{align}
where $m_1$ and $m_2$ are the primary and secondary stars of a binary
star, $F_2/F_1$ is the flux ratio of the secondary star to the primary
star, $a_1$ is the angular semi-major axis of the primary star, $K_1$
is the semi-amplitude of the radial velocity of the primary star, and
$\varpi$, $P$, $e$, and $i$ are the parallax, period, eccentricity,
and inclination angle of the binary star, respectively. We define a
primary star as a star observed by astrometry and spectroscopy, and a
secondary star as a fainter star than the primary star. The secondary
star is an unseen star if $F_2/F_1=0$. We can get $a_1$, $\varpi$,
$P$, $e$, and $i$ from astrometry, and $K_1$ from spectroscopy. We
have to remark that $\fmspectro$ is similar to but different from the
spectroscopic mass function ordinarily defined (hereafter
$\fmspectroconv$), since we obtain $\fmspectro$, dividing
$\fmspectroconv$ by $\sin^3 i$. We can know the inclination angle,
$i$, thanks to astrometric observation, and thus mainly refer to
$\fmspectro$, not $\fmspectroconv$.

Practically, we calculate $\fmspectro$ of {\tt AstroSpectroSB1} binary
stars as
\begin{align}
  \fmspectro = \left[ \left( \frac{C_1}{{\rm au}} \right)^2 + \left(
    \frac{H_1}{{\rm au}} \right)^2 \right]^{3/2} \left( \frac{P}{{\rm
      yr}} \right)^{-2} \sin^{-3} i \; [\msun],
\end{align}
where $C_1$ and $H_1$ are Thiele-Innes elements
\citep{1960pdss.book.....B, 1978GAM....15.....H}, derived by
spectroscopic observation. On the other hand, we calculate
$\fmspectro$ of {\tt Orbital} binary stars, substituting half {\tt
  rv\_amplitude\_robust} into $K_1$.

We regard binary stars as BH binary candidates if they satisfy the
following two conditions:
\begin{align}
  &0.5 \le \fmspectro/\fmastro \le
  2, \label{eq:ConditionOfSimilarMassFunctions} \\
  &\fmastro \ge 3 \msun. \label{eq:ConditionOfHighMassFunction}
\end{align}
We adopt the first condition expressed by
Eq. (\ref{eq:ConditionOfSimilarMassFunctions}) for the following
reason. When a binary star is a BH binary, the secondary star is an
unseen star; $F_2/F_1=0$. Substituting $F_2/F_1=0$ into
Eq. (\ref{eq:fmastro1}), we find $\fmastro=\fmspectro$. Thus, BH
binaries should satisfy $\fmastro \simeq \fmspectro$.  By the second
condition of Eq. (\ref{eq:ConditionOfHighMassFunction}), we can select
binary star candidates with $m_2 \ge 3\msun$ irrespective of
$m_1$. Such binary stars are likely to be BH binaries, since the
maximum mass of neutron stars is expected to be $\sim 2 \msun$
\citep{1996ApJ...470L..61K}.

\begin{figure}[ht!]
  \plotone{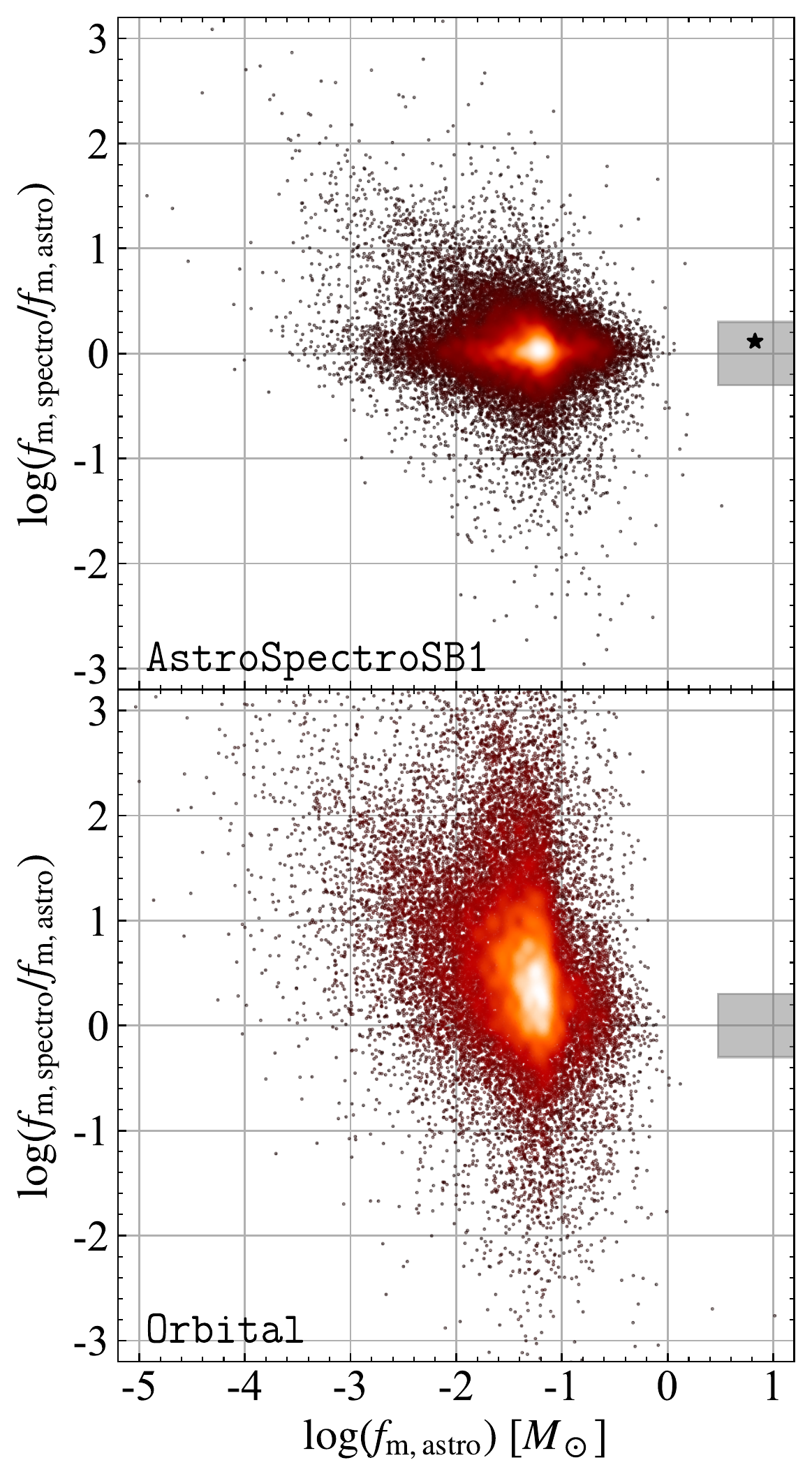}
  \caption{Scatter plots of $\fmastro$ and $\fmspectro/\fmastro$ for
    {\tt AstroSpectroSB1} (top) and {\tt Orbital} (bottom) binary
    stars. The color scale represents the square root of the relative
    density of binary stars. Shaded regions satisfy the two conditions
    of BH binary candidates expressed as
    Eqs. (\ref{eq:ConditionOfSimilarMassFunctions}) and
    (\ref{eq:ConditionOfHighMassFunction}). The BH binary candidate
    found in this work (GDR3 5870569352746779008) is
    emphasized as a star in the top panel.}
  \label{fig:searchForBhCandidate}
\end{figure}

Figure \ref{fig:searchForBhCandidate} shows $\fmastro$ and
$\fmspectro/\fmastro$ of all the samples.  The shaded region in this
figure corresponds to the two conditions imposed in this study
(Eqs. (\ref{eq:ConditionOfSimilarMassFunctions}) and
(\ref{eq:ConditionOfHighMassFunction})).  Only one binary star
satisfies these two conditions.  Its basic parameters are summarized
in Table \ref{tab:BasicParameters}. We analyze this BH binary
candidate in later sections.

In general, we have $\fmspectro \ge \fmastro$ for any binary stars,
which can be easily confirmed from their definitions in
Eq. (\ref{eq:fmastro1}) and Eq. (\ref{eq:fmspectro1}).  However,
Figure \ref{fig:searchForBhCandidate} shows that the distribution of
$\fmspectro/\fmastro$ spreads under 1. There can be two
reasons. First, $\fmspectro$ is underestimated for the second type of
binary stars. For these binary stars, we adopt {\tt
  rv\_amplitude\_robust} for $K_1$ in
Eq. (\ref{eq:fmspectro2}). However, the observed radial velocities may
not fall at the right phase to fully sample the orbit's maximum and
minimum radial velocities. Second, some of binary stars contain large
errors of either $\fmspectro$ or $\fmastro$, while they have
$\fmspectro \ge \fmastro$ in reality. In fact, such binary stars may
hide BH binaries. However, in this paper, we conservatively select
binary stars with $\fmspectro \simeq \fmastro$ as BH binary
candidates.  This is because the small discrepancy between
$\fmspectro$ and $\fmastro$ is anticipated for a binary system in
which the secondary star is much fainter than the primary star
($F_2/F_1 \simeq 0$).

It is a bit strange that the $\log (\fmspectro/\fmastro)$ values are
centered on zero for both the {\tt AstroSpectroSB1} and {\tt Orbital}
binary stars. Typically, binary stars should have luminous secondary
stars \citep[e.g.][]{2012Sci...337..444S}, and thus should have
$F_2/F_1$ close to 1, and large $\fmspectro/\fmastro$ (or small
$\fmastro$). The reason for this discrepancy might be that {\it Gaia}
preferentially select binary stars with faint secondary stars.

\begin{figure}[ht!]
  \plotone{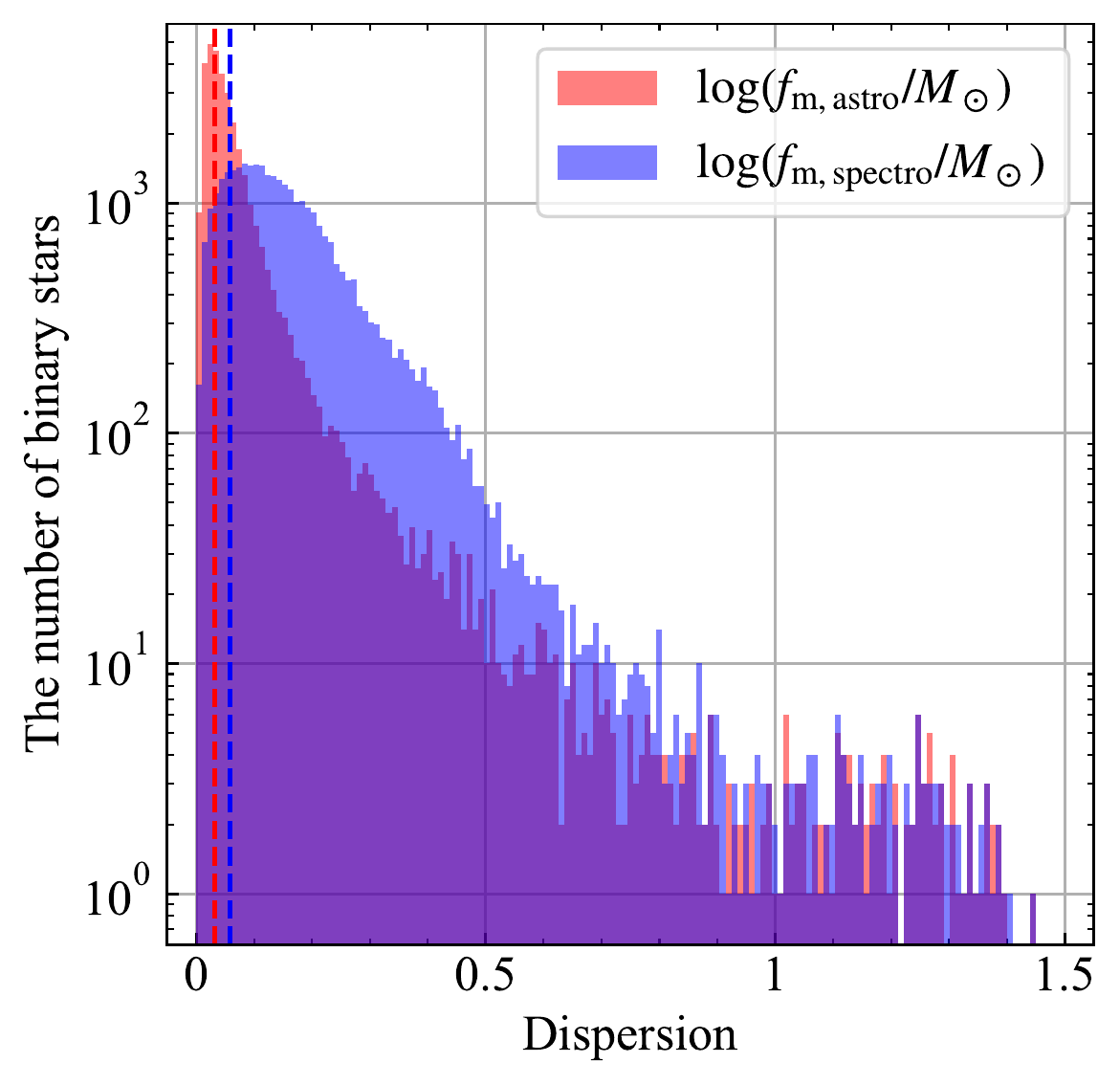}
  \caption{Distributions of the $\log \fmastro$ and $\log \fmspectro$
    dispersions for {\tt AstroSpectroSB1} binary stars. The red and
    blue dashed lines indicate the $\log \fmastro$ and $\log
    \fmspectro$ dispersions, respectively, for the BH binary candidate
    (GDR3 source ID 5870569352746779008). }
  \label{fig:readAndViewErrorOfAstrospectrosb1}
\end{figure}

Figure \ref{fig:readAndViewErrorOfAstrospectrosb1} shows the
distributions of the $\log \fmastro$ and $\log \fmspectro$ dispersions
for {\tt AstroSpectroSB1} binary stars. In order to obtain these
dispersions, we generate $10^3$ Monte Carlo random draws of the
covariance matrix of the BH binary candidate in the GDR3 {\tt
  nss\_two\_body\_orbit} table. Note that the number of Monte Carlo
random draws is sufficient, since the distributions for $10^3$ random
draws are similar to those for $10^2$ random draws.  We cannot
calculate the $\log \fmspectro$ dispersions of {\tt Orbital} binary
stars, since the covariance matrix does not include the data of {\tt
  rv\_amplitude\_robust}. The $\log \fmspectro$ dispersions of {\tt
  Orbital} binary stars should be larger than that of {\tt
  AstroSpectroSB1} binary stars. This is also the reason why $\log
(\fmspectro/\fmastro)$ values spread more widely in {\tt Orbital}
binary stars than in {\tt AstroSpectroSB1} binary stars. The peak of
the $\log \fmspectro$ dispersion is at $\sim 0.2$. On the other hand,
$\gtrsim 100$ binary stars have the $\log \fmspectro$ dispersions of
$\gtrsim 1$. This should affect the presence of binary stars with
$\log \fmspectro/\fmastro < 0$.

If a binary star system is actually a triple star system, $\fmspectro$
will be significantly overestimated, and consequently $\fmspectro >
\fmastro$ for the following reason. Astrometric observation is
sensitive to the outer binary's motion: the relative motion between
the inner binary and third star. On the other hand, spectroscopic
observation is sensitive to the inner binary's motion, since its
motion velocity is much larger than the outer binary's motion
velocity. Thus, we will calculate $\fmspectro$ in
Eq. (\ref{eq:fmspectro2}), using $P$, $e$, and $i$ of the outer binary
and $K_1$ of the inner binary. This $\fmspectro$ will be larger than
the actual $\fmspectro$ of both the inner and outer binaries. For
obtaining the inner (outer) binary's $\fmspectro$, the adopted $P$
($K_1$) is larger than the actual inner (outer) binary's. This may
happen in the {\tt Orbital} binary stars more frequently.

\subsection{Some comments on rejected binaries}
\label{sec:SomeComments}

Before analyzing the BH binary candidate in detail, we review our
search. In particular, we focus on binary stars which look like BH
binaries at a glance, but which our search rejects. GDR3 provides the
{\tt binary\_masses} table including the masses of primary and
secondary stars estimated from the PARSEC isochrone
models\footnote{\url{http://stev.oapd.inaf.it/cgi-bin/cmd}}
\citep{2012MNRAS.427..127B}. We can obtain such binary stars with
following ADQL query:
\begin{verbatim}
select nss.*, gs.*, bm.*
from gaiadr3.nss_two_body_orbit as nss,
gaiadr3.gaia_source as gs,
gaiadr3.binary_masses as bm
where gs.source_id = nss.source_id
and bm.source_id = nss.source_id
and (nss.nss_solution_type = 'AstroSpectroSB1'
	 or (nss.nss_solution_type = 'Orbital'
         and gs.rv_amplitude_robust IS NOT NULL)
	 or (gs.non_single_star = 3))
\end{verbatim}
We just add the {\tt binary\_masses} table to the ADQL query in
section \ref{sec:SearchForBH}. Note that our samples are the ones
obtained with the ADQL query in section \ref{sec:SearchForBH} unless
otherwise stated. Not all our samples are listed in the {\tt
  binary\_masses} table, because the mass estimation is only applied
to primary stars in the main sequence (MS) on the color-magnitude
diagram. In the {\tt binary\_masses} table, there are 6 {\tt
  AstroSpectroSB1} and 3 {\tt Orbital} binary stars containing
secondary stars with $>3 \msun$. In spite of their secondary masses,
none of them are regarded as BH binary candidates by our search.

As for the 6 {\tt AstroSpectroSB1} binary stars, they are rejected,
because all of them have too large $\fmspectro/\fmastro$ ($>10$). This
means that, although these binary stars have main-sequence primary
stars with $1$--$2$ $\msun$, they have secondary stars with $>3$
$\msun$ and smaller (but non-zero) luminosity than the primary
stars. It is difficult to interpret these binary stars as BH
binaries. Thus, we remove them from our list of BH binary candidates.
The 3 {\tt Orbital} binary stars are ruled out, since they have too
small $\fmspectro/\fmastro$ ($<0.01$). Incomprehensibly, their
$F_2/F_1$ values are negative. Astrometric or spectroscopic results
might not be appropriate. In fact, all of them have large
goodness-of-fit values ($>5$), where the goodness-of-fit is expected
to obey the normal distribution if astrometric parameters are
correctly derived. When \cite{2022arXiv220700680A} search for NS and
BH binaries, they rule out binary stars with goodness-of-fit values
more than 5 from NS and BH binary candidates.

The second condition expressed by
Eq. (\ref{eq:ConditionOfHighMassFunction}) may be too strict to
complete a search for BH binaries from our sample. This condition
means that the secondary mass is more than $3$ $\msun$ for any primary
masses. We convert this condition to $m_2>3$ $\msun$, where $m_2$ is
drawn from the lower limit of $m_2$ ({\tt m2\_lower}) in the GDR3 {\tt
  binary\_masses} table. By this conversion, we can relax our search
for BH binaries, since the secondary mass can be more than $3$ $\msun$
even for $\fmastro<3$ $\msun$ if the primary mass is larger than a
certain value. However, we find no other BH binary candidate. Although
the two conditions expressed by
Eqs. (\ref{eq:ConditionOfSimilarMassFunctions}) and
(\ref{eq:ConditionOfHighMassFunction}) are slightly strict, we confirm
that there is only one BH binary candidate (GDR3 source ID
5870569352746779008) in GDR3 astrometric binary stars with
spectroscopic data.

\section{Analysis of a BH candidate}
\label{sec:Analysis}

We summarize the basic parameters of the BH binary candidate in Table
\ref{tab:BasicParameters}. For the right ascension, declination, BP-RP
color, reddening of BP-RP color, [M/H], and surface gravity ($\log
g$), we adopt the mean values in the GDR3 {\tt gaia\_source}
table. The galactic longitude and latitude are derived from the right
ascension and declination. We obtain the mean value of the extinction
in G band ($A_{\rm G}$) from the EXPLORE G-Tomo scientific data
application \citep{2022A&A...661A.147L,
  2022A&A...664A.174V}\footnote{\url{https://explore-platform.eu/}},
while the value in the parentheses is the mean value of the GDR3 {\tt
  gaia\_source} table. Hereafter, we adopt the former value for the
extinction. We obtain the goodness-of-fit value from the GDR3 {\tt
  nss\_two\_body\_orbit} table. In order to calculate the mean values
and one standard deviation intervals of the distance, period ($P$),
physical semi-major axis ($a_1/\varpi$), eccentricity ($e$),
inclination ($i$), radial velocity semi-amplitude ($K_1$), astrometric
mass function ($\fmastro$), and spectroscopic mass function
($\fmspectro$), we generate $10^4$ Monte Carlo random draws of the
covariance matrix of the BH binary candidate in the GDR3 {\tt
  nss\_two\_body\_orbit} table\footnote{The number of Monte Carlo
random draws is sufficiently large, since the results are similar if
we adopt $10^3$ for the number of the random draws.}. In this method,
we also obtain $\fmastro>5.68$ $\msun$ and
$\fmspectro>6.57$ $\msun$ at a probability of $99$ \%. Note that
the distance is calculated from the parallax in the GDR3 {\tt
  nss\_two\_body\_orbit} table, not in the GDR3 {\tt gaia\_source}
table. According to \cite{2022arXiv220605595G}, the parallax in the
former table is more accurate than in the latter table. We get the
absolution magnitude in G band ($M_{\rm G}$) from the mean of apparent
magnitude in the GDR3 {\tt gaia\_source}, and the mean of the distance
derived above.

\begin{deluxetable*}{ll}
  \tablecaption{Basic parameters of the BH binary
    candidate. \label{tab:BasicParameters}}
  \tablehead{Quantities & Values}
  \startdata
  (1) Source ID & 5870569352746779008 \\
  (2) Orbital solution   & {\tt AstroSpectroSB1} \\
  (3) Right ascension    & $207.5697^\circ$ \\
  (4) Declination        & $-59.2390^\circ$ \\
  (5) Galactic longitude & $310.4031^\circ$ \\
  (6) Galactic latitude  & $2.7765^\circ$ \\
  (7) Absolute magnitude in G band ($M_{\rm G}$) & $1.95$ mag \\
  (8) Extinction in G band ($A_{\rm G}$)         & $0.5628$ mag ($0.70$ mag) \\
  (9) BP-RP color                                & $1.49$ mag \\
  (10) Reddening of BP-RP color                  & $0.37$ mag \\
  (11) Surface gravity ($\log g$, {\tt logg\_gspphot}) & $3.25$ [cgs] \\
  (12) $[{\rm M/H}]$ ({\tt mh\_gspphot}) & $0.0066$ dex \\
  (13) Goodness-of-Fit            & $3.07$ \\
  \hline
  (14) Distance                   & $1164.41 \pm 25.16$ pc \\
  (15) Period ($P$)               & $1352.22 \pm 45.81$ day \\
  (16) Physical semi-major axis ($a_1/\varpi$)  & $4.5194 \pm 0.1305$ au \\
  (17) Eccentricity ($e$)         & $0.5323 \pm 0.0153$ \\
  (18) Inclination ($i$)          & $35.15 \pm 0.99^\circ$ \\
  (19) Radial velocity semi-amplitude ($K_1$)     & $27.0 \pm 1.0$ km s$^{-1}$ \\
  (20) Astrometric mass function ($\fmastro$)     & $6.75 \pm 0.51$ $\msun$ \\
  (21) Lower bound in $\fmastro$ (99\%)   & $\fmastro > 5.68 \msun$ \\
  (22) Spectroscopic mass function ($\fmspectro$) & $8.85 \pm 1.13$ $\msun$ \\
  (23) Lower bound in $\fmspectro$ (99\%) & $\fmspectro > 6.57 \msun$ \\
  (24) Probability of $\fmspectro>\fmastro$ & $95$ \% \\
  (25) Probability of $\fmspectro<\fmastro$ & $ 5$ \%
  \enddata
  \tablecomments{ From row 3 (right ascension) to 13 (goodness-of-fit)
    except for row 8 (Extinction in G band), we show the mean
    value in GDR3. For row 8 (Extinction in G band), the value is
      obtained from the EXPLORE G-Tomo scientific data application,
      and the value in the parentheses is obtained from the mean value
      in GDR3. From row 14 (distance) to 20 ($\fmastro$) as well as
    in row 22 ($\fmspectro$), we show the mean value and one standard
    deviation interval.  In rows 21 and 23, we show the 99\%
    confidence level of $\fmastro$ and $\fmspectro$, respectively.  In
    rows 24 and 25, we show the probabilities of $\fmspectro >
    \fmastro$ and $\fmspectro<\fmastro$, respectively (see section
    \ref{sec:Analysis} for more detail).}
\end{deluxetable*}

The goodness-of-fit value, $3.07$, is relatively low, since
\cite{2022arXiv220700680A} consider that NS and BH binary candidates
should have the goodness-of-fit value less than $5$. Note that the
goodness-of-fit value for reliable sources should be normally
distributed with a mean of zero. Thus, we are not going to arguing
that the BH binary candidate is reliable only from the goodness-of-fit
value. Nevertheless, we have to remark that the goodness-of-fit value
is typical of the {\tt AstroSpectroSB1} binary stars as described
later (see Figure \ref{fig:drawGoodnessOfFit}). Although the
goodness-of-fit value largely deviates from zero, it would not
directly mean that this BH binary candidate is unreliable. We find
that the ratios of means to standard deviation intervals are high for
$\fmastro$ and $\fmspectro$ ($13.2$ and $7.83$, respectively). They
should be relatively well-measured. Additionally, the $\log \fmastro$
and $\log \fmspectro$ dispersions are small, compared to those of
other {\tt AstroSpectroSB1} binary stars, as seen in Figure
\ref{fig:readAndViewErrorOfAstrospectrosb1}. This should be another
evidence that the parameters of this BH candidate are
well-measured. Moreover, at a probability of $99$ \%, $\fmastro>5.68$
$\msun$ and $\fmspectro>6.75$ $\msun$.  These values are unlikely to
fall below $3$ $\msun$. A concern is that $\fmspectro$ is
systematically larger than $\fmastro$, which we discuss in section
\ref{sec:Discussion}.

\begin{figure}[ht!]
  \plotone{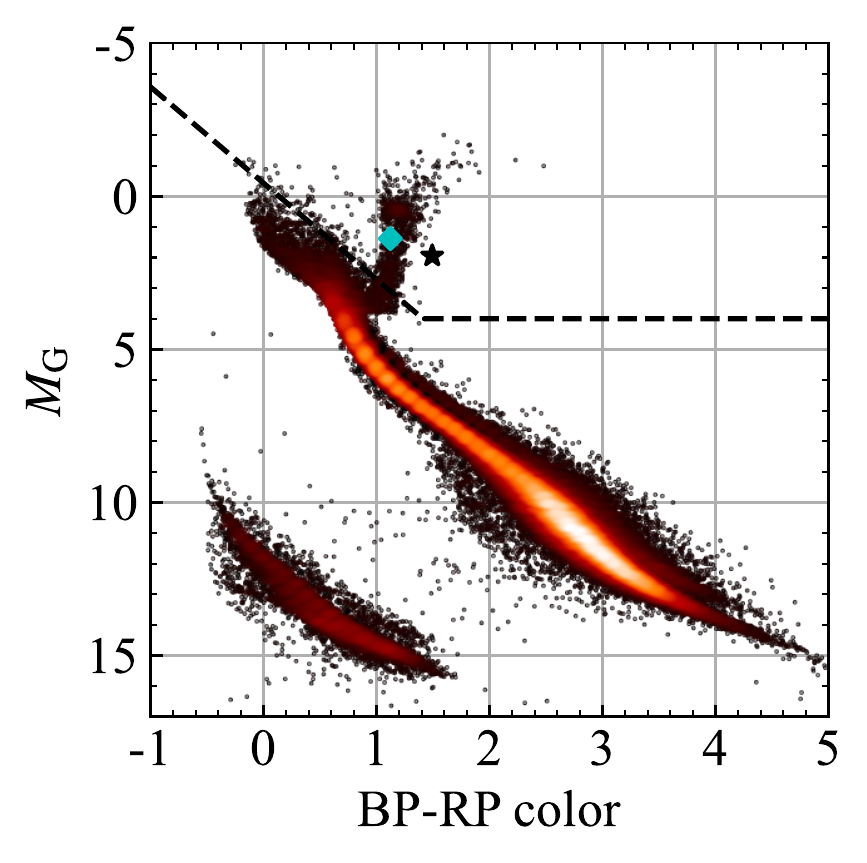}
  \caption{Color magnitude diagram of stars in the GDR3 {\tt
      gaia\_source} table. These stars are filtered in the same way as
    those in figure 6c of \cite{2018A&A...616A..10G}. The color scale
    represents the square root of the relative density of stars. The
    star and diamond points indicate the BH binary candidate (GDR3
    source ID 5870569352746779008), where the star and diamond points
    are not corrected and corrected by its extinction and reddening,
    respectively. We define the regions of MS and RGB stars below and
    above the dashed line, respectively. The line is expressed as
    Eq. (\ref{eq:MS-RGBBoundary}).}
  \label{fig:drawHrDiagram}
\end{figure}

Figure \ref{fig:drawHrDiagram} shows the color-magnitude diagram of
the primary star of the BH binary candidate, and GDR3 stars whose
G-band absolute magnitudes and BP-RP colors are well-measured. MS and
red giant branch (RGB) regions are defined as regions below and above
the dashed line. The dashed line is expressed as
\begin{align}
  M_{\rm G} = \left\{
    \begin{array}{ll}
      3.14 \left( {\rm BP} - {\rm RP} \right) - 0.43 & ( {\rm BP} -
      {\rm RP}<1.41) \\
      4 & ({\rm otherwise})
    \end{array}
    \right.. \label{eq:MS-RGBBoundary}
\end{align}
The first case of Eq. (\ref{eq:MS-RGBBoundary}) is the same as in
\cite{2022arXiv220604648A}. We induce the second case to avoid
regarding low-mass MS stars as RGB stars. As seen in Figure
\ref{fig:drawHrDiagram}, the primary star of the BH binary candidate
is likely to be a RGB star. This is consistent with its small surface
gravity ($\log g = 3.25$). The primary star indicated by the star
point (not corrected by its extinction and reddening) is redder than
RGB stars on the color-magnitude diagram. It suffers from interstellar
reddening, since it is located in the Galactic disk ($b =
2.7765^\circ$).  In fact, the primary star indicated by the diamond
point (corrected by its extinction and reddening) is on the RGB.

Generally, BH binary candidates are thought dubious when their primary
stars are RGB stars. This is because such primary stars can easily
outshine companion stars even if the companion stars are more massive
than the primary stars. Moreover, it is difficult to estimate the
masses of RGB stars in binary systems. Such RGB stars can be in
so-called Algol-type systems \citep{2022MNRAS.512.5620E}. They can be
luminous but low-mass (say $\sim 0.1$ $\msun$) if they experience mass
transfer. These types of problems frequently happen in BH binary
candidates with only spectroscopic data, or usual spectroscopic mass
function, $\fmspectroconv$ (not $\fmspectro$), and $\fmspectroconv$ is
$\sim 1$ $\msun$. In order to conclude that their secondary stars are
$>3$ $\msun$ compact objects, we need to estimate the primary stars'
masses and inclination angles of the binary stars.  As an
illustration, let us consider a spectroscopic binary characterized by
$\fmspectroconv = 1$ $\msun$ and inclination angle $i=60^\circ$ (from
which we obtain $\fmspectro=1.54$ $\msun$).  If the the primary star's
mass is $1.2$ $\msun$, the secondary star's mass is $3$ $\msun$.  In
this case, the $3$ $\msun$ secondary star is highly likely a BH.
In contrast, if the the primary star's mass is $0.2$ $\msun$, the
secondary star's mass is $1.9$ $\msun$.  In this case, we cannot
exclude the possibility that the $1.9$ $\msun$ secondary star is a
main-sequence star outshined by the primary RGB star.  From this
simple illustrative example, we can see that the mass estimation of
primary stars critically affects whether their secondary stars are BH
or not.

Fortunately, these types of problems do not happen in our BH binary
candidate. We know the inclination angle $i$ of the binary star from
the astrometric data, and get $\fmspectro$ in a model-independent
way. Moreover, this BH binary candidate has $\fmastro>5.68$ $\msun$
and $\fmspectro>6.75$ $\msun$ at a probability of 99 \%. {\it The
  secondary mass is more than $5$ $\msun$, even if this BH binary
  candidate is an Algol-type system, or the primary RGB mass is close
  to zero.} The primary RGB star cannot outshine the $>5$ $\msun$
secondary star even if the secondary star is in the main-sequence
phase, or the faintest among $5$ $\msun$ stars in any phases except a
BH. This point is described in detail below. Thus, the secondary star
is likely to be a BH.

We examine the possibility that the secondary star of the BH binary
candidate may be a single object except a BH, or multiple star
systems. When a stellar mass is fixed, a MS star is the faintest
except stellar remnants like WD, NS, and BH. If a MS star with the
same mass as the secondary star is more luminous than the primary
star, the possibility that the secondary star is a single object
except a BH can be ruled out. When the total mass of a multiple star
system is fixed, a multiple star system with equal-mass MS stars is
the least luminous. This is because MS stars become luminous more
steeply with their masses increasing. If an $n$-tuple star system with
equal-mass MS stars has the same mass as the secondary star, and
larger luminosity than the primary star, the possibility that the
secondary star is any $n$-tuple star systems can be rejected. Thus, we
compare the luminosity of the primary star with the luminosities of a
single MS star or multiple MS star systems with equal masses.

\begin{figure}[ht!]
  \plotone{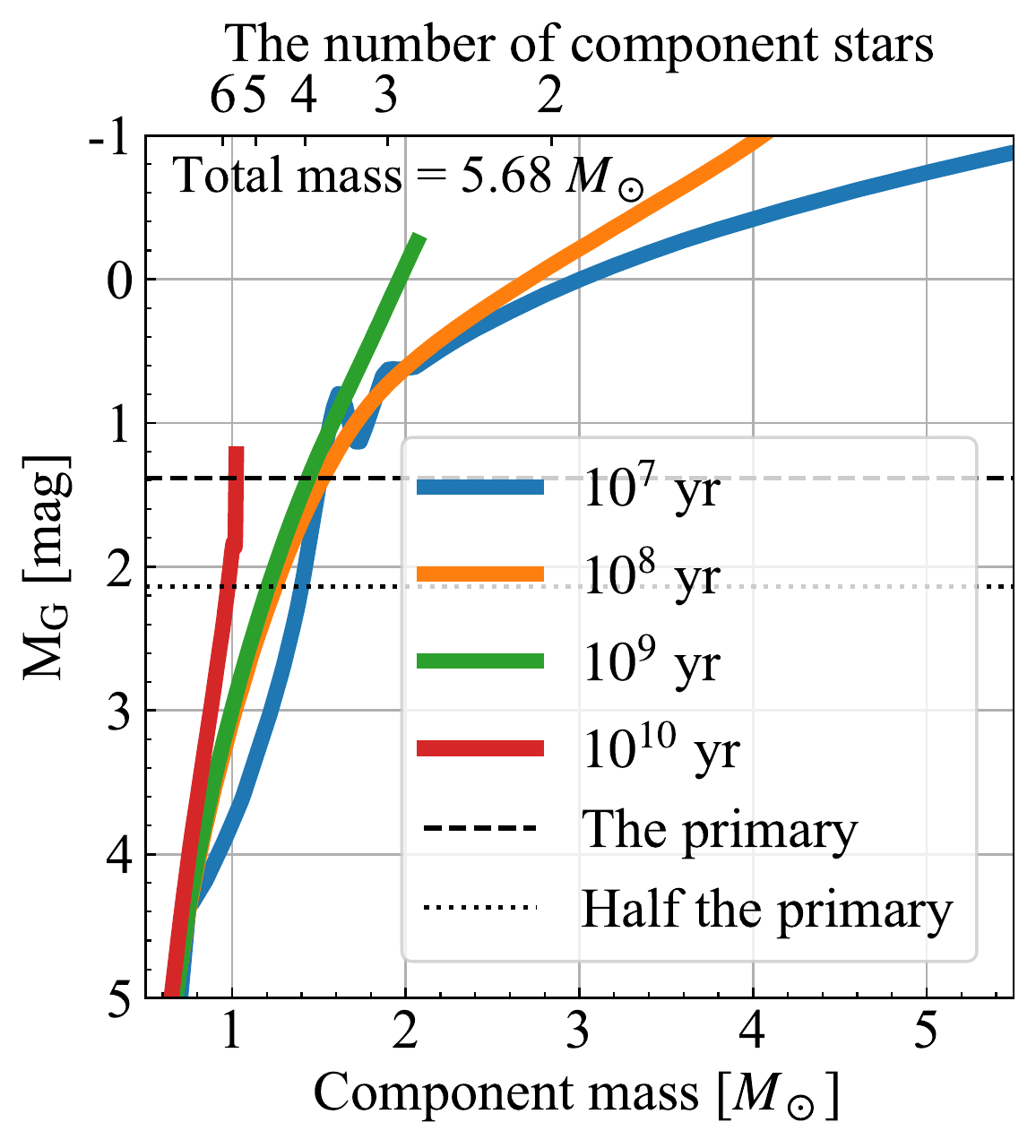}
  \caption{G-band absolute magnitude of multiple star systems with
    equal-mass MS stars whose ages are $10^7$, $10^8$, $10^9$, and
    $10^{10}$ yrs. The total mass of the multiple star systems is
    $5.68$ $\msun$, the lower bound mass of the secondary star of the
    BH binary candidate at a probability of 99 \%. The component mass
    and the number of stars are shown in the lower and upper $x$-axis,
    respectively.  We show only MS stars defined in
    Eq. (\ref{eq:MS-RGBBoundary}). That is the reason why the curves
    of $10^9$ and $10^{10}$ yrs cut off in the middle.  We obtain
    G-band absolute magnitude and BP-RP color at each mass and age,
    using the PARSEC code \citep{2012MNRAS.427..127B}. The metallicity
    is set to the solar metallicity, the same as the primary star of
    the BH binary candidate. It seems that there is no publicly
    available spectroscopic survey data that provides reliable
    metallicity for the primary star of our BH binary candidate.  The
    dashed line indicates the G-band absolute magnitude of the primary
    star, which is corrected by the G-band extinction. The dotted line
    indicates the G-band absolute magnitude of a star half as luminous
    as the primary star.}
  \label{fig:estimateComponentMass}
\end{figure}

Figure \ref{fig:estimateComponentMass} shows the G-band absolute
magnitude of multiple star systems with equal-mass MS stars. The total
mass of the multiple star systems is $5.68$ $\msun$, the lower bound
mass of the secondary star of the BH binary candidate at a probability
of 99 \%. We can rule out single, binary and triple stars with the
total mass of $5.68$ $\msun$. They would outshine the primary star if
they were the secondary star. A quadruple star system with each
stellar mass of $1.4$--$1.5$ $\msun$ is as luminous as the primary
star. However, such a quadruple star system should be detected by {\it
  Gaia} itself. A quintuple star system with each stellar mass of
$1.1$ $\msun$ has luminosity twice less than the primary star, and
might not be observed by {\it Gaia}. Except for multiple star systems
with MS stars, the secondary star can be a triple NS star system or a
quadruple WD star system, where the maximum mass of NS and WD are
about $2.0$ and $1.4$ $\msun$, respectively. Such systems may be more
valuable than a single BH, since they have never been discovered to
our knowledge. In any case, the secondary star should be quadruple or
higher-order star systems except for a single BH. Moreover, the size
of the system should be more compact than the pericenter distance of
the primary star, $\sim 2.4$ au. It is unclear that such multiple
systems are stable under the perturbation of the primary star.

\begin{figure}[ht!]
  \plotone{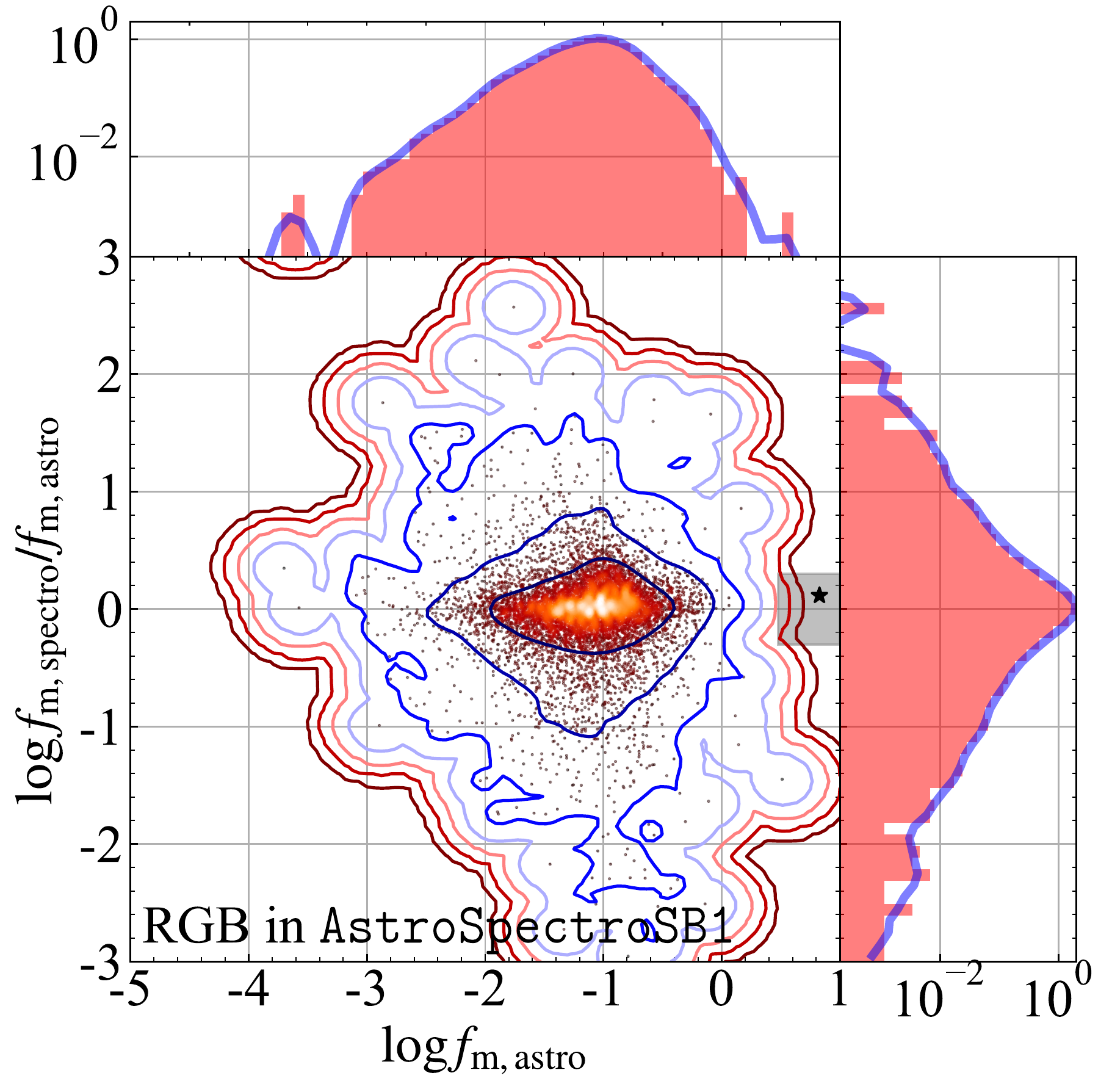}
  \caption{Bottom left: Scatter plots of $\fmastro$ and
    $\fmspectro/\fmastro$ for RGB stars in {\tt AstroSpectroSB1}. The
    color scale represents the square root of the relative density of
    binary stars. Contours indicate $\sigma$ levels of $1$, $2$,
    $\cdots$, and $7$ from the inner to the outer. The shaded region
    is considered to calculate the p-values in Table
    \ref{tab:P-values}. The p-values are calculated by a
    kernel-density estimate with the kernel bandwidth of Scott's rule
    \citep{1992mde..book.....S}. The star point indicates the BH
    candidate (GDR3 source ID 5870569352746779008). It is not included
    in the samples with which the p-values are calculated. Top and
    left: $\fmastro$ and $\fmspectro/\fmastro$ distributions,
    respectively. The histograms indicate the sample distribution, and
    the curves indicate the projected distributions derived by the
    kernel-density estimate.}
  \label{fig:calcPValueOfMassFunction}
\end{figure}

In order to assess whether the BH binary candidate is coincidentally
located on the $\fmastro$--$\fmastro/\fmspectro$ plane, we calculate
the p-values of a $\fmastro$--$\fmastro/\fmspectro$ region around the
BH binary candidate. We adopt a kernel-density estimate with a kernel
bandwidth of Scott's rule \citep{1992mde..book.....S}. The bandwidth
is $N_{\rm sample}^{-1/6}$, where $N_{\rm sample}$ is the number of
samples. At first, we select RGB primary stars from {\tt
  AstroSpectroSB1} as samples for the kernel-density estimate. The
number of samples is $9047$. Note that the BH binary candidate is
excluded from the samples. Figure \ref{fig:calcPValueOfMassFunction}
shows the kernel-density contours of $1$, $2$, $\cdots$, and $7\sigma$
levels from the inner to the outer. We calculate the p-value in the
shaded region. The p-value is $9.6 \times 10^{-12}$, and the $\sigma$
level is $6.1$. The position of $\fmastro$ and $\fmastro/\fmspectro$
of the BH binary candidate is unlikely to be coincident.

\begin{deluxetable*}{lllll}
  \tablecaption{P-values. \label{tab:P-values}}
  \tablehead{Sample & Number & p-value & $\sigma$ level & Remark}
  \startdata
  All & $64095$ & $2.4 \times 10^{-12}$ & $7.0$ & \\
  All in {\tt AstroSpectroSB1}       & $33466$ & $9.1 \times 10^{-12}$ & $6.8$ & \\
  Low-error in {\tt AstroSpectroSB1} & $28188$ & $1.0 \times 10^{-11}$ & $6.8$ & Exclude samples with top 10 \% large errors \\
  Low-error in {\tt AstroSpectroSB1} & $17614$ & $1.1 \times 10^{-11}$ & $6.8$ & Exclude samples with errors more than 0.2 in log-scale \\
  RGBs in {\tt AstroSpectroSB1}           & $9047$ & $9.6 \times 10^{-12}$ & $6.1$ & The same samples used in Figure \ref{fig:calcPValueOfMassFunction} \\
  Low-error RGBs in {\tt AstroSpectroSB1} & $8626$ & $7.5 \times 10^{-10}$ & $6.2$ & Exclude samples with top 10 \% large errors \\
  Low-error RGBs in {\tt AstroSpectroSB1} & $5395$ & $1.2 \times 10^{-9}$  & $6.1$ & Exclude samples with errors more than 0.2 in log-scale
  \enddata
\end{deluxetable*}

We select samples for the kernel-density estimate in different ways in
order to investigate whether the p-values depend on the choice of
samples. We summarize the choices of samples and their results in
Table \ref{tab:P-values}. The first column indicates the choice of
samples. Note that the BH binary candidate is not included in any
choices. For ``All'', we choose all the samples selected in section
\ref{sec:SampleSelection}. For ``All in {\tt AstroSpectroSB1}'', we
choose all the samples in {\tt AstroSpectroSB1}. For ``RGBs in {\tt
  AstroSpectroSB1}'', we extract only the RGB primary stars in the
samples of ``All in {\tt AstroSpectroSB1}''. This is the samples shown
in Figure \ref{fig:calcPValueOfMassFunction}. We also make samples,
excluding samples with large errors of $\fmastro$ and $\fmspectro$
from ``All in {\tt AstroSpectroSB1}'' and ``RGBs in {\tt
  AstroSpectroSB1}''. We calculate the errors in the same way as the
one standard deviation of the BH binary candidate in Table
\ref{tab:BasicParameters}, where we generate $10^3$ Monte Carlo random
draws for each sample for calculation cost savings. We adopt two cases
to exclude samples. In the first case, we exclude 10 \% samples with
largest errors in either of $\fmastro$ and $\fmspectro$. In the second
case, we exclude samples with errors larger than $0.2$ in log-scale
for either of $\fmastro$ and $\fmspectro$. Note that $0.2$ is similar
to the bandwidth of the kernel-density estimate. In any cases, the
p-values are small, and the $\sigma$ levels are high. The position of
$\fmastro$ and $\fmastro/\fmspectro$ of the BH binary candidate is
unlikely to be coincident, independently of the choices of samples for
the kernel-density estimate.

We search for the BH binary candidate in several database. The GDR3
{\tt variability} table \citep{2022arXiv220606416E} and the All-Sky
Automated Survey for SuperNovae
\citep[ASAS-SN;][]{2017PASP..129j4502K} do not include the BH binary
candidate as a variable star. Its light curve is available on the
ASAS-SN Photometry Database \citep{2014ApJ...788...48S,
  2019MNRAS.485..961J}\footnote{\url{https://asas-sn.osu.edu/photometry}. The
BH binary candidate is observed in V and g band over $\sim 3000$
days. We do not find any periodic feature. The Transiting Exoplanet
Survey Satellite \citep[TESS;][]{2015JATIS...1a4003R} has performed
twice high-cadence observations during about 30 days according to data
downloaded from
TESScut\footnote{\url{https://mast.stsci.edu/tesscut/}} for the BH
binary candidate. The duration is too short to detect its periodic
variability due to its binary orbit if any, since it has a period of
about 1000 days. Wide-field Infrared Survey Explorer
\citep[WISE;][]{2010AJ....140.1868W} also observes the BH binary
candidate over $\sim 4000$ days according to ALLWISE Multiepoch
Photometry
Table\footnote{\url{https://irsa.ipac.caltech.edu/cgi-bin/Gator/nph-scan?submit=Select&projshort=WISE}}
and NEOWISE-R Single Exposure (L1b) Source
Table\footnote{\url{https://irsa.ipac.caltech.edu/cgi-bin/Gator/nph-dd}}. We
do not recognize any periodic variability. At the time of September
2022,} the BH binary candidate is not listed in the following data
base: SIMBAD\footnote{\url{http://simbad.cds.unistra.fr/simbad/}}, the
ninth catalog of spectroscopic binary orbits
\citep[SB9;][]{2004A&A...424..727P}, RAdial Velocity Experiments
\citep[RAVE;][]{2017AJ....153...75K}, the Galactic Archaeology with
HERMES \citep[GALAH;][]{2021MNRAS.506..150B}, the Large sky Area
Multi-Object fiber Spectroscopic Surveys
\citep[LAMOST;][]{2012RAA....12.1197C}, and the Apache Point
Observatory Galactic Evolution Experiment
\citep[APOGEE;][]{2017AJ....154...94M}. High-energy telescopes, such
as the Fermi gamma-ray space telescope \citep{2009ApJ...697.1071A},
the Swift Burst Alert Telescope \citep[Swift
  BAT:][]{2005SSRv..120..143B} XMM-Newton \citep{2001A&A...365L..18S},
the Chandra observatory \citep{2000SPIE.4012....2W}, and the Galaxy
Evolution Explorer \citep[GALEX;][]{2005ApJ...619L...1M}, do not
observe it as far as we see Aladin
lite\footnote{\url{https://aladin.u-strasbg.fr/AladinLite/}}. ESO
archive\footnote{\url{http://archive.eso.org/scienceportal/home}} does
not list it. In summary, we do not find any positive nor negative
evidence for the BH binary candidate.

\section{Discussion}
\label{sec:Discussion}

First, we compare the BH binary candidate with other BH binary
candidates by previous studies, and assess whether our BH binary
candidate is similar to others rejected before. As described in
section \ref{sec:Analysis}, BH binary candidates tend to be rejected
when their primary stars are RGB stars. It is difficult to estimate
the masses of RGB stars, and such binary stars can be Algol-type
systems in which primary stars are low-mass (say $\sim 0.1$
$\msun$). Since such BH binary candidates have $\fmspectroconv \sim 1$
$\msun$, the mass estimate of RGB stars severely affect the secondary
mass. However, our BH binary candidate has $\fmastro > 5.68$ $\msun$
and $\fmspectro > 6.75$ $\msun$ at a probability of 99 \%. In this
case, the secondary mass is more than $\sim 5$ $\msun$ even if the
primary mass is nearly zero. Note that the secondary mass increases
monotonically with the primary mass increasing when $\fmastro$ or
$\fmspectro$ is fixed. Thus, the secondary star is likely to be a BH,
even if the BH binary candidate is an Algol-type system.

\cite{2022arXiv220605595G} listed up BH binary candidates with $\sim
2$ $\msun$ MS stars and $\sim 3$ $\msun$ BHs. However,
\cite{2022MNRAS.515.1266E} pointed out possibility that they are
Algol-type systems consisting of $\sim 0.2$ $\msun$ stripped stars and
$\sim 2$ $\msun$ MS stars. The reason for this discrepancy is as
follows. \cite{2022arXiv220605595G} thought that $\sim 2$ $\msun$ MS
stars dominate the luminosity (photometry) and radial-velocity motion
(spectroscopy) of the binary stars. On the other hand,
\cite{2022MNRAS.515.1266E} claimed that $\sim 2$ $\msun$ MS stars
dominate the luminosity, while $\sim 0.2$ $\msun$ stripped stars
dominate the radial-velocity motion. This interpretation better
explains their spectral energy distribution and spectroscopic mass
function ($\fmspectroconv \sim 1.5$ $\msun$) more naturally. We do not
expect that similar things happen in our BH binary candidate for the
following reason. If a hidden star dominates the radial-velocity
motion, we replace $m_2$ with $m_1$ in
Eq. (\ref{eq:fmspectro1}). Since $\fmastro \sim \fmspectro$, we obtain
$m_1=4 \fmastro (1+F_2/F_1)^2$ $\msun$ and $m_2= 4 \fmastro
(1+F_2/F_1)^2(1+2F_2/F_1)$ $\msun$. Thus, the RGB primary mass should
be at least $4 \fmastro$ ($\sim 23$) $\msun$. However, its luminosity
requires its mass much less than $23$ $\msun$. Thus, a hidden star
does not dominate the radial-velocity motion of our BH binary
candidate in contrary to the BH binary candidates in table 10 of
\cite{2022arXiv220605595G}.

\cite{2022arXiv220605595G} also show another table of BH binary
candidates (their table 9) in which BH binary candidates belong to
{\tt SB1}, and have high $\fmspectroconv$ ($>3$ $\msun$). Hereafter,
we call them ``Gaia's table 9 candidates''. Although these candidates
have secondary stars with more than $3$ $\msun$ for any primary
masses, \cite{2022arXiv220605595G} cannot rule out that the secondary
stars consist of multiple star systems, similarly to our description
in section \ref{sec:Analysis}. We remark that our BH candidate will be
better-constrained than all of Gaia's table 9 candidates. Our BH
binary candidate has larger mass function and smaller luminosity than
Gaia's table 9 candidates except for GDR3 source IDs
4661290764764683776 and 5863544023161862144. GDR3 source ID
4661290764764683776 has high $\fmspectroconv$ ($=13.67$ $\msun$),
however its primary star has high luminosity, $-6.707$ mag in G
band. Since the primary star can be more luminous than a $\sim 13$
$\msun$ MS star, it is difficult to confirm that the secondary star is
a BH. GDR3 source ID 5863544023161862144 shows eclipses, and
consequently its secondary should not be a BH
\citep{2022arXiv220605595G}. In summary, we can easiest rule out the
possibility that the secondary star of our BH candidate consists of a
multiple star system.

\begin{figure*}[ht!]
  \plotone{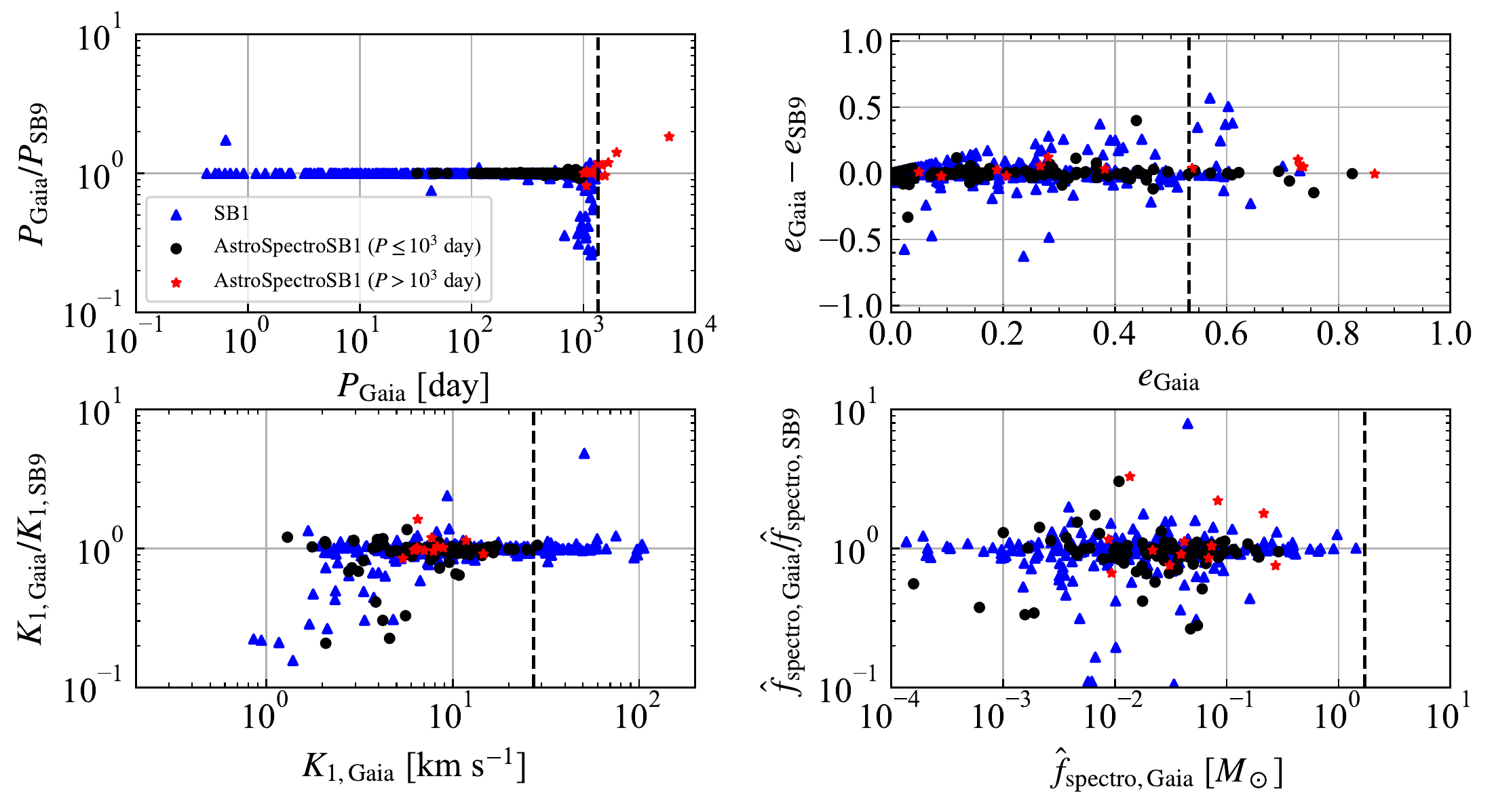}
  \caption{Comparison of orbital parameters (Period, eccentricity,
    radial velocity semi-amplitude, and $\fmspectroconv$) between GDR3
    and SB9. Triangle, circle and star points indicate GDR3 {\tt SB1},
    {\tt AstroSpectroSB1} with a period of $\le 10^3$ day, and {\tt
      AstroSpectroSB1} with a period of $> 10^3$ day,
    respectively. The dashed lines show the mean values of the orbital
    parameters of the BH binary candidate (GDR3 source ID
    5870569352746779008).}
  \label{fig:crossMatchingGaiaSb9}
\end{figure*}

\cite{2022gdr3.reptE...7P} and \cite{2022arXiv220705086J} compared
orbital parameters in GDR3 with those in SB9
\citep{2004A&A...424..727P}, in particular for spectroscopic binary
stars with either one component being parameterized ({\tt SB1}). They
found that Gaia's and SB9's periods are inconsistent for periods of
more than $10^3$ days in SB9. Since they did not investigate {\tt
  AstroSpectroSB1}, we investigate both of {\tt SB1} and {\tt
  AstroSpectroSB1}. We find 304 {\tt SB1} and 109 {\tt
  AstroSpectroSB1} in common between GDR3 and SB9. Our BH binary
candidate is not included in SB9 as described in the previous section.
In Figure \ref{fig:crossMatchingGaiaSb9}, we make comparison between
orbital parameters of binary stars in GDR3 and SB9. Note that the
$x$-axes in Figure \ref{fig:crossMatchingGaiaSb9} adopt GDR3 values,
while \cite{2022gdr3.reptE...7P} and \cite{2022arXiv220705086J} adopt
SB9 values for the $x$-axes in their figure 7.41 and figure 6,
respectively. Similarly to \cite{2022gdr3.reptE...7P} and
\cite{2022arXiv220705086J}, we find that periods in GDR3 are largely
different from those in SB9 for {\tt SB1} with periods of more than
$10^3$ days. However, for {\tt AstroSpectroSB1}, their periods do not
deviate up to periods of a few $10^3$ days. The other parameters in
GDR3 are also in good agreement with those in SB9 for {\tt
  AstroSpectroSB1}, in particular around the mean values of the
orbital parameters of the BH binary candidate. This does not directly
show that the spectroscopic data of the BH binary candidate is
reliable, since most of binary stars in SB9 are brighter than our BH
binary candidate. Nevertheless, this means that GDR3 values of {\tt
  AstroSpectroSB1} binary stars may be reliable even if the binary
stars have periods of a few $10^3$ days.

\cite{2022MNRAS.517.3888B} compared GDR3 {\tt SB1} with the database
of LAMOST \citep{2012RAA....12.1197C} and GALAH
\citep{2021MNRAS.506..150B}, and found that GDR3 {\tt SB1} with
periods of less than $10^{1.5}$ days may be refuted. Although our BH
binary candidate belongs to {\tt AstroSpectroSB1} (not to {\tt SB1}),
it has a period of $\gtrsim 10^3$ days, much larger than $10^{1.5}$
days. Our BH binary candidate may not be refuted by the criteria of
\cite{2022MNRAS.517.3888B}.

\cite{2022arXiv220700680A} and \cite{2023MNRAS.518.2991S}
independently presented lists of NS and BH binary candidates in
GDR3. Their lists do not include our BH binary candidate. This is
because they focus on binary stars with primary MS stars. The masses
of MS stars can be estimated less model-dependently than those of RGB
stars. The masses and natures of secondary stars can be derived
robustly. Thus, they avoided binary stars with primary RGB stars. On
the other hand, although the primary star of our BH binary candidate
is a RGB star, we can call it a ``BH binary candidate'', because its
$\fmastro$ and $\fmspectro$ are high; $\gtrsim 5.68$ and $6.75$
$\msun$, respectively, at a probability of 99 \%. Its secondary mass
is more than $5$ $\msun$, regardless of the primary RGB mass.

\begin{figure}[ht!]
  \plotone{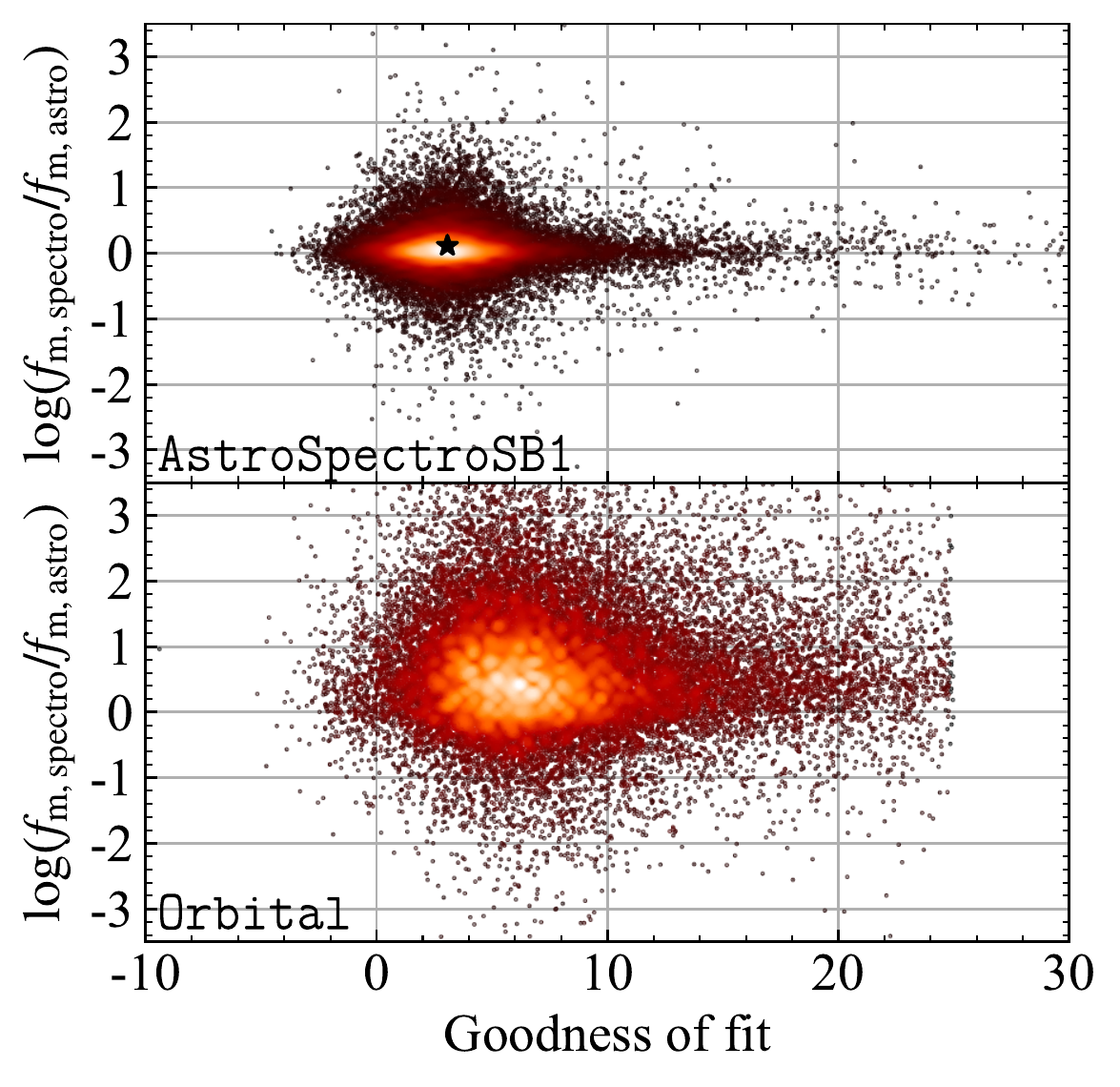}
  \caption{The ratio of $\fmspectro$ to $\fmastro$ as a function of
    goodness-of-fit values for {\tt AstroSpectroSB1} (top) and {\tt
      Orbital} (bottom) binary stars. The color scale represents the
    square root of the relative density of binary stars. The star point
    indicates the BH binary candidate (GDR3 source ID
    5870569352746779008).}
  \label{fig:drawGoodnessOfFit}
\end{figure}

Conversely, we examine the lists of \cite{2022arXiv220700680A} and
\cite{2023MNRAS.518.2991S} from our conditions. We focus on binary
stars with $m_2>2$ $\msun$ in their lists. Note that the maximum NS
mass can be $\sim 2$ $\msun$. Our sample selected in section
\ref{sec:SampleSelection} does not include BH binary candidate in the
Andrews's list. The candidates do not have spectroscopic data. This
may be partly because astrometric binary stars with spectroscopic data
(i.e. our sample) has systematically large goodness-of-fit values.
Figure \ref{fig:drawGoodnessOfFit} shows that goodness-of-fit values
in {\tt AstroSpectroSB1} and {\tt Orbital} are centered at $\sim 3$
and $\sim 5$, respectively. Actually, this can be seen in the middle
panel of figure 4 in \cite{2022arXiv220700680A}. Their figure 4
includes all the {\tt Orbital} binary stars with and without
spectroscopic data, and indicates the second peak around the
goodness-of-fit value of $\sim 5$. The second peak should consist of
{\tt Orbital} binary stars with spectroscopic data. We do not know the
reason for the systematic upward shift. We have to remark that bright
binary stars (G band magnitude of $<13$), i.e. those with
spectroscopic data, have the systematically higher goodness-of-fit
values, while faint binary stars (G band magnitude of $>13$)
typically have lower goodness-of-fit values\footnote{The exact
  value of $13$ is obtained by \cite{2023arXiv230207880E}.}. In any
case, our sample does not include the list of
\cite{2022arXiv220700680A}, because they avoid including binary stars
with goodness-of-fit values of more than 5 in their list.

Our sample includes Shanaf's three BH binary candidates (GDR3 source
IDs: 3263804373319076480, 3509370326763016704, and
6281177228434199296). However, we do not list up them as BH binary
candidates. This is because their $\fmspectro/\fmastro$ are small
($0.25$, $0.0053$, $0.0017$, respectively) for our first condition as
seen in Eq. (\ref{eq:ConditionOfSimilarMassFunctions}). We do not
intend to reject the three BH candidates completely. The three BH
candidates may suffer from large errors of spectroscopic data, and
consequently have small $\fmspectro/\fmastro$. We suspect this
possibility, because two of the three BH candidates are not included
in {\tt AstroSpectroSB1} binary stars despite the fact that they have
spectroscopic data. Our sample selected in section
\ref{sec:SampleSelection} does not include the other 5 BH binary
candidates because of the absence of spectroscopic data.

A few days after we posted this work on arXiv,
\cite{2023MNRAS.518.1057E} reported one promising BH binary candidate
different from our BH binary candidate. They made follow-up
spectroscopic observation, and showed that {\it Gaia}'s astrometric
data is consistent with their spectroscopic data. Since their BH
binary candidate has a shorter period ($185.6$ days) than our binary
candidate ($1352.22$ days), they can finish their follow-up
observation in a short period of time. They also mentioned our BH
binary candidate, and did not conclude whether our BH binary candidate
is genuine because of the absence of follow-up spectroscopic
observations. Their argument is in good agreement with ours. Note that
we analyze our BH binary candidate in detail.

Several BH binary candidates can be rejected for exceptional
reasons. Although \cite{2022arXiv220605595G} found that GDR3 source ID
2006840790676091776 has high $\fmspectroconv$, they did not include it
in their list of BH binary candidates. This is because it is close to
a bright star, whose apparent magnitude is 3.86 mag in G band. There
is no such bright stars close to our BH binary candidate. Nearby stars
have apparent magnitude of at least 13 mag in G band. The reason for
this rejection can not be applied to our BH binary
candidate. \cite{2022arXiv220700680A} removed GDR3 source ID
4373465352415301632\footnote{This object is later confirmed as a BH
binary (also known as Gaia BH1) by \cite{2023MNRAS.518.1057E}.}, since
its period ($\sim 186$ days) is roughly 3 times {\it Gaia}'s scanning
period ($63$ days). Our BH binary candidate has a period of $1352$
days, not integer multiple of {\it Gaia}'s scanning period.

\begin{figure}[ht!]
  \plotone{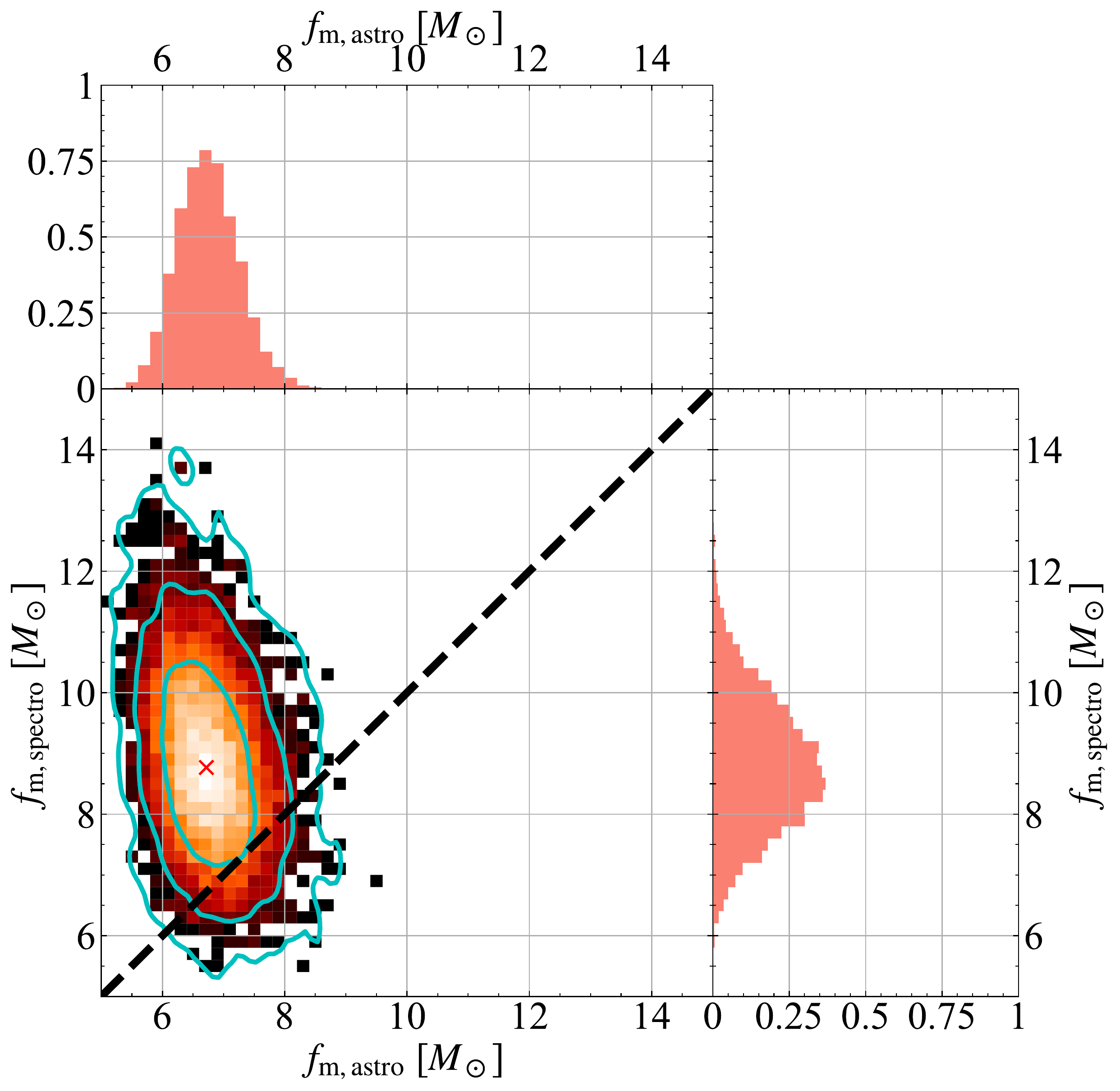}
  \caption{Bottom left: 2-dimensional probability distribution of
    $\fmastro$ and $\fmspectro$ for the BH binary candidate (GDR3
    source ID 5870569352746779008). The cross point means the typical
    values of $\fmastro$ and $\fmspectro$. Contours indicate $\sigma$
    levels of $1$, $2$, and $3$ from the inner to the outer. The
    dashed line shows $\fmastro = \fmspectro$. Top and left:
    $\fmastro$ and $\fmspectro$ distributions, respectively.}
  \label{fig:calcBasicParameter_massfunctions}
\end{figure}

Hereafter, we make several concerns. First of all, we mostly rely on
GDR3 astrometric and spectroscopic data, which are already largely
processed. We do not assess correctness of the data of our BH binary
candidate. Aside from this, we find that the BH binary candidate has
$\fmspectro>\fmastro$ and $\fmspectro<\fmastro$ at probabilities of
$95$ and $5$ \%, respectively (see Table
\ref{tab:BasicParameters}). Although $\fmspectro=\fmastro$ is
possible, $\fmspectro$ is always larger than $\fmastro$ in the
$1\sigma$-level region as seen in Figure
\ref{fig:calcBasicParameter_massfunctions}. For comparison, we
calculate the probabilities of $\fmspectro>\fmastro$ and
$\fmspectro<\fmastro$ for GDR3 source ID 5136025521527939072 which is
in {\tt AstroSpectroSB1}, and suggested as a NS binary candidate by
\cite{2022arXiv220605595G}\footnote{The NS binary candidate is more
likely to be a WD binary according to
\cite{2023MNRAS.518.1057E}.}. They are $72$ and $28$ \%. Both of
$\fmspectro>\fmastro$ and $\fmspectro<\fmastro$ are in the
$1\sigma$-level region, in contrast to our BH binary candidate. The
$\fmspectro$ and $\fmastro$ of our BH binary candidate are not as
similar as those of GDR3 source ID 5136025521527939072. Nevertheless,
we may regard $\fmspectro=\fmastro$, since our BH binary candidate may
contain some systematic error in either of spectroscopic or
astrometric data.

Another concern is that the primary star of the BH binary candidate is
a RGB star. Theoretical studies \citep[e.g.][]{2020PASJ...72...45S,
  2022ApJ...928...13S} expected that a BH binary with a $\gtrsim 10$
$\msun$ MS primary star is likely to be found first \citep[but see
  also][]{2023arXiv230107207S}. This is because such MS stars are
bright, and can be observed even if they are distant. Moreover, they
are longer-lived than RGB stars with similar masses. However, GDR3
does not present orbital parameters of binary stars with $\gtrsim 10$
$\msun$ MS primary stars in {\tt AstrospectroSB1} nor {\tt Orbital}
according to the GDR3 {\tt binary\_masses} table obtained with the
ADQL query in section \ref{sec:SomeComments}. We do not know the
reason for the absence of such binary stars in GDR3. Nevertheless,
when there are no such binary stars, it may be natural that a BH
binary with a RGB star is first discovered.

We need two types of follow-up observations in order to assess if the
BH binary candidate is true or not. The first type should be
spectroscopic observations to verify the GDR3 spectroscopic data, and
to perform spectral disentangling of the BH binary candidate similar
to \cite{2022MNRAS.515.1266E}. The second type should be deep
photometric observations. Such observations could constrain whether
the secondary star is a BH, or consists of multiple stars. We remark
that \cite{2023arXiv230207880E} have carried out these follow-up
observations, and confirmed it as a genuine BH binary, Gaia BH2. This
demonstrates that these follow-up observations would be important for
confirming or refuting future BH candidates, which may be discovered
by our search methodology in upcoming Gaia data.

\section{Summary}
\label{sec:Summary}

We first search for BH binary candidates from astrometric binary stars
with spectroscopic data in GDR3. From the sample of 64108 binary
stars, we find one BH binary candidate. The GDR3 source ID is
5870569352746779008. Since its primary star is a RGB star, we cannot
estimate the mass of the primary RGB star. However, because of its
high astrometric and spectroscopic mass function ($\fmastro > 5.68$
$\msun$ and $\fmspectro > 6.75$ $\msun$ at a probability of 99 \%),
the secondary star should have more than $5$ $\msun$, and is likely to
be a BH, regardless of the primary mass. Unless the secondary star is
a BH, it must be quadruple or higher-order multiple star systems with
the total mass of $5.68$ $\msun$. To rule out the possibility of
multiple star systems, we need deep photometric observations. Rather,
if it is quadruple or higher-order multiple star systems, long-term
observation may find modulation of the primary's orbit
\citep[e.g.][]{2020ApJ...890..112H, 2020ApJ...897...29H,
  2022PhRvD.106l3010L}.

\begin{figure}[ht!]
  \plotone{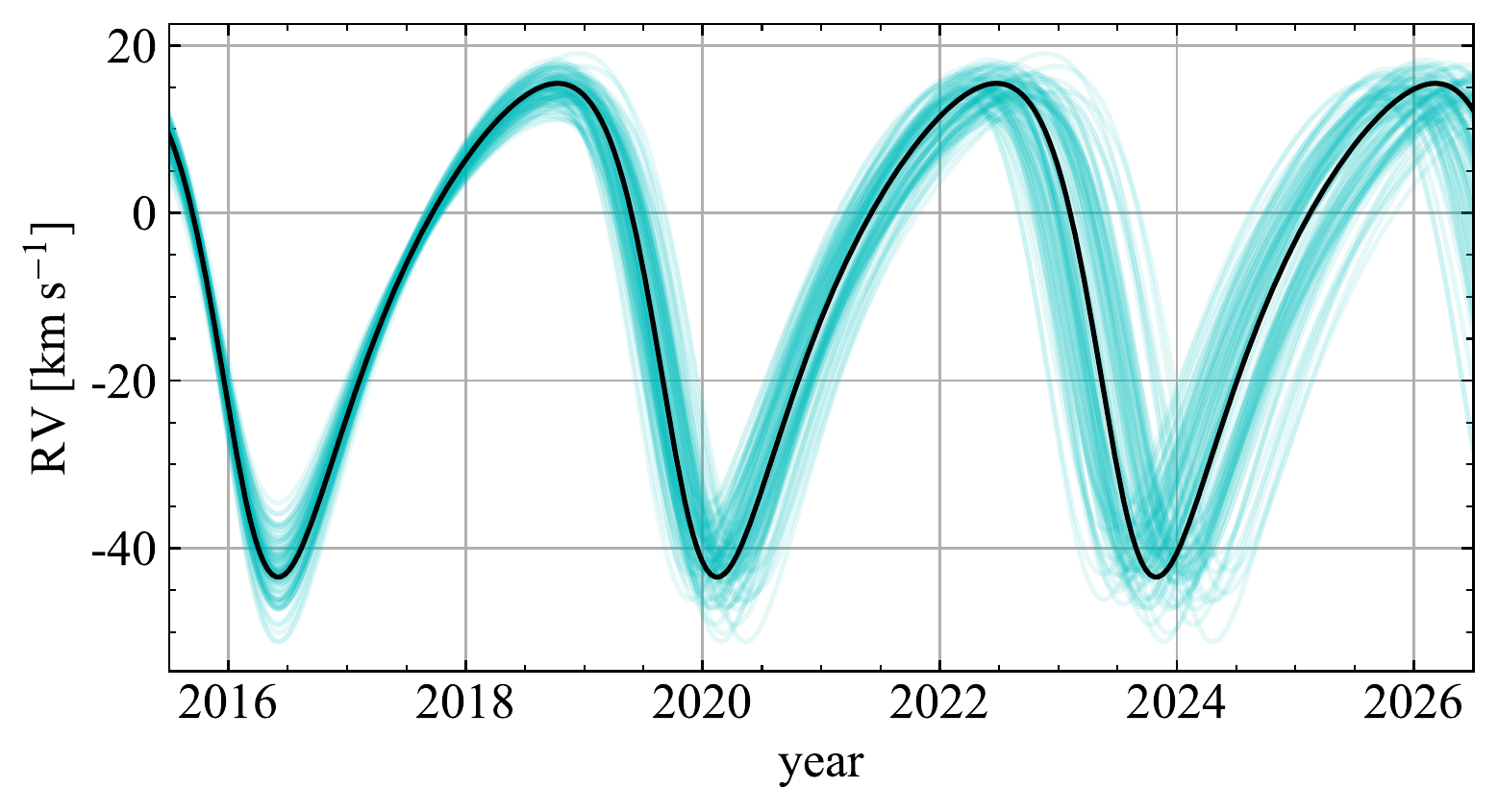}
  \caption{Predicted radial velocities of the BH binary candidate
    (GDR3 source ID 5136025521527939072), based on GDR3. The black
    curve can be obtained from typical values in GDR3. We generate
    cyan curves from $10^2$ Monte Carlo random draws, using the
    covariance matrix. The radial velocity amplitude is larger and the
    turnover time spreads more widely than those in the upper panel of
    figure 3 in \citep{2023arXiv230207880E}, because we do not adopt
    any explicit link between the spectroscopic and astrometric
    Thiele-Innes parameters.}
  \label{fig:reconstructOrbit_rv}
\end{figure}

The weakness of this paper is that our conclusion entirely relies on
the Gaia DR3. In particular, our BH binary candidate has a period of
$\sim 1300$ days, more than a period of 34 months of the Gaia DR3 data
collection. Our conclusion have to be confirmed by follow-up
observations. For example, we need the time evolution of the radial
velocity of our BH binary candidate similarly to that of Gaia BH1
obtained by \cite{2023MNRAS.518.1057E}. Figure
\ref{fig:reconstructOrbit_rv} shows the predicted radial velocities of
the BH binary candidate. Eventually, \cite{2023arXiv230207880E} have
confirmed the radial velocity variability, by observing it $\sim 30$
times from the last half of 2022 to the beginning of 2023, when the
radial velocities had steeply decreased.

Previously, RGB stars harboring BHs have not been searched for because
of the difficulty of the mass estimation of the RGB stars (and thus
BHs).  However, our tentative discovery in this paper encourages us to
explore not only BHs orbiting around MS stars but also BHs orbiting
around RGB stars in the future data releases of {\it Gaia}.

\section*{Acknowledgments}
  We thank the anonymous referee for many fruitful advices.  This
  research could not been accomplished without the support by
  Grants-in-Aid for Scientific Research (17H06360, 19K03907) from the
  Japan Society for the Promotion of Science.  KH is supported by JSPS
  KAKENHI Grant Numbers JP21K13965 and JP21H00053.  NK is supported by
  JSPS KAKENHI Grant Numbers JP22K03686.  TK is supported by JSPD
  KAKENHI Grant Number JP21K13915, and JP22K03630.  MS acknowledges
  support by Research Fellowships of Japan Society for the Promotion
  of Science for Young Scientists, by Forefront Physics and
  Mathematics Program to Drive Transformation (FoPM), a World-leading
  Innovative Graduate Study (WINGS) Program, the University of Tokyo,
  and by JSPS Overseas Challenge Program for Young Researchers.

  This work presents results from the European Space Agency (ESA)
  space mission Gaia. Gaia data are being processed by the Gaia Data
  Processing and Analysis Consortium (DPAC). Funding for the DPAC is
  provided by national institutions, in particular the institutions
  participating in the Gaia MultiLateral Agreement (MLA). The Gaia
  mission website is \url{https://www.cosmos.esa.int/gaia}. The Gaia archive
  website is \url{https://archives.esac.esa.int/gaia}.

\software{{\tt Matplotlib} \citep{2007CSE.....9...90H}; {\tt NumPy}
  \citep{2011CSE....13b..22V}; {\tt Astropy}
  \citep{2013A&A...558A..33A}; {\tt SciPy}
  \citep{2020NatMe..17..261V}}


\end{document}